\documentclass[prb,amsmath,amssymb,twocolumn,showpacs,floatfix]{revtex4}
\usepackage{graphicx}
\usepackage[english]{babel}
\usepackage{times}
\usepackage{dcolumn}
\usepackage[usenames,dvipsnames]{color}
\usepackage{units}

\begin{document}

\title{Quantitative determination of the discretization and truncation
errors in the numerical renormalization-group calculations of spectral
functions}

\author{Rok \v{Z}itko}

\affiliation{Jo\v{z}ef Stefan Institute, Jamova 39, SI-1000 Ljubljana,
Slovenia,\\
Faculty  of Mathematics and Physics, University of Ljubljana,
Jadranska 19, SI-1000 Ljubljana, Slovenia}

\date{\today}

\begin{abstract}
In the numerical renormalization group (NRG) calculations of spectral
functions of quantum impurity models, the results are always affected
by discretization and truncation errors. The discretization errors can
be alleviated by averaging over different discretization meshes
(``z-averaging''), but since each partial calculation is performed for
a finite discrete system, there are always some residual
discretization and finite-size errors. The truncation errors affect
the energies of the states and result in the displacement of the delta
peak spectral contributions from their correct positions. The two
types of errors are interrelated: for coarser discretization, the
discretization errors increase, but the truncation errors decrease
since the separation of energy scales is enhanced. In this work, it is
shown that by calculating a series of spectral functions for a range
of the total number of states kept in the NRG truncation, it is
possible to estimate the errors and determine the error-bars for
spectral functions, which is important when making accurate comparison
to the results obtained by other methods and for determining the
errors in the extracted quantities (such as peak positions, heights,
and widths). The closely related problem of spectral broadening is
also discussed: it is shown that the overbroadening contorts the
results without, surprisingly, reducing the variance of the curves. It
is thus important to determine the results in the limit of zero
broadening. The method is applied to determine the error bounds for
the Kondo peak splitting in external magnetic field. For moderately
large fields, the results are consistent with the Bethe Ansatz study
by Moore and Wen. We also discuss the regime of large $U/\Gamma$
ratio. It is shown that in the high-field limit, a spectral step is
observed in the spectrum precisely at the Zeeman frequency until the
field becomes so large that the step merges with the atomic spectral
peak.
\end{abstract}


\maketitle

\newcommand{\vc}[1]{{\mathbf{#1}}}
\newcommand{\vck}{\vc{k}}
\newcommand{\braket}[2]{\langle#1|#2\rangle}
\newcommand{\expv}[1]{\langle #1 \rangle}
\newcommand{\corr}[1]{\langle\langle #1 \rangle\rangle}
\newcommand{\ket}[1]{| #1 \rangle}
\newcommand{\Tr}{\mathrm{Tr}}
\newcommand{\kor}[1]{\langle\langle #1 \rangle\rangle}
\newcommand{\degg}{^\circ}
\renewcommand{\Im}{\mathrm{Im}}
\renewcommand{\Re}{\mathrm{Re}}
\newcommand{\dtN}{{\dot N}}
\newcommand{\dtQ}{{\dot Q}}
\newcommand{\GG}{{\mathcal{G}}}
\newcommand{\atanh}{\mathrm{atanh}}
\newcommand{\sgn}{\mathrm{sgn}}

\section{Introduction}

Quantum impurity physics \cite{hewson, bulla2008} is an active area of
research, which is particularly important for the problems of magnetic
properties of confined electrons (quantum dots \cite{kouwenhoven2001,
andergassen2010}, magnetic impurity atoms on surfaces
\cite{ternes2009, brune2009}), but also for strongly correlated
electron systems due to the mapping of bulk correlated models to
self-consistent single-impurity models within the dynamical mean-field
theory \cite{georges1996, kotliar2006}. For strong electron-electron
interactions, the quantum impurity models are notoriously difficult to
solve and they generally require the application of non-perturbative
techniques. One such technique is the numerical renormalization group
(NRG) \cite{wilson1975, krishna1980a, bulla2008}, which consists of
discretizing the continuum of the conduction-band states, transforming
the problem to the form of a semi-infinite tight-binding chain, and
numerically diagonalizing the resulting discrete Hamiltonian in an
iterative way. The discretization is performed by splitting the energy
band into intervals of widths that decrease as a geometric series
($\propto \Lambda^{-n}$, where $\Lambda>1$ is called the
discretization parameter) as the Fermi level is approached
\cite{wilson1975}. This particular choice of the discretization scheme
is adapted to the behavior of the Kondo model, where each energy scale
makes a comparable contribution to the renormalization group flow of
the exchange coupling constant in the scaling regime
\cite{wilson1975,anderson1970}. The NRG was first applied to calculate
the thermodynamic properties of impurity problems \cite{wilson1975,
krishna1975, krishna1980a, krishna1980b, cragg1980, oliveira1981}, and
was later extended to dynamical properties
\cite{oliveira1981phaseshift, frota1986, sakai1989, costi1994}.
Further important improvements were the development of the
density-matrix approach to spectral function calculation
\cite{hofstetter2000}, the self-energy trick \cite{bulla1998}, and the
introduction of the complete-Fock-space basis \cite{anders2005,
anders2006, peters2006, weichselbaum2007} which solved the
overcounting problem.

Since the calculation is performed for a discretized problem, one
expects significant systematic discretization errors. They appear, for
example, in the form of oscillations in the calculated spectral
functions with frequency $\log\Lambda$ and its harmonics (in
logarithmic frequency space). These oscillations can be reduced by
performing the so-called $z$-averaging, wherein one performs the same
NRG calculation for several interleaved discretization meshes and
averages the results \cite{yoshida1990, campo2005, resolution}. By
averaging over two meshes, one cancels out $\log\Lambda$ oscillations,
by averaging over four meshes one cancels out $2\log\Lambda$
oscillators, etc., thus the $z$-averaging is best performed for $N_z$
meshes where $N_z=2^n$. Using an improved discretization scheme
\cite{resolution}, the cancellation of oscillations is remarkable even
in the case of strong hybridization of the impurity with the
conduction band states and for large values of the discretization
parameter $\Lambda$. Nevertheless, the spectral functions calculated
using the NRG are always affected by the discretization and the
finite-size errors to some degree, even when all technical refinements
are used \cite{resolution}.

Another source of systematic errors in the NRG is the truncation.
Since the Fock space grows exponentially with the chain length (by a
factor of 4 for a spinfull single-channel impurity problem), the set
of states kept after each step is truncated to some finite number $N$.
This is clearly an approximation, which was, however, shown to lead to
highly accurate results \cite{wilson1975}. The approach works because
of the ``energy-scale-separation'' property of quantum impurity
problems: the matrix elements coupling high-energy and low-energy
excitations are small and controlled by $\Lambda$, large $\Lambda$
leading to stronger decoupling.

Finally, for particle-hole asymmetric baths there is a further source
of error in the NRG (the ``mass-flow effect'') due to the iterative
algorithm used in the NRG for integrating out the impurity bath
degrees of freedom, since the impurity parameter shift due to the real
part of the bath propagator at a given NRG step only includes the
contribution from the chain sites already included in the calculation,
while the the contribution of the remaining half-infinite chain is
missing \cite{vojta2010}. The mass-flow effect is particularly
severe for bosonic baths, while for fermionic baths it was found
to have little effect \cite{vojta2010}.

While the presence of the systematic errors in the NRG calculations is
common knowledge \cite{wilson1975, bulla2008, vojta2010}, few
systematic studies have actually appeared in the literature and the
dependence of the errors on the calculation parameters is still not
widely known.
The purpose of this work is to analyze the discretization and
truncation errors in the spectral functions obtained in the most
sophisticated calculations using the complete Fock-space (CFS)
approach \cite{peters2006,weichselbaum2007} with the self-energy trick
\cite{bulla1998} and the improved discretization mesh with averaging
over many values of the $z$-parameter \cite{resolution}. It will be
shown that with the increasing number $N$ of states kept, the spectral
function obtained in the CFS approach exhibits variations which arise
due to the discrete nature of the NRG Hamiltonian and the particular
way of collecting the spectral information in the CFS technique
\cite{peters2006}, which is susceptible to truncation errors. By
calculating the statistical properties of these variations, one can
obtain useful information which quantifies the unavoidable
discretization and truncation errors in the NRG. The error estimates
thus obtained are actually lower error bounds, since it is conceivable
that in addition to the errors which lead to variations as a function
of the NRG calculation parameters ($N$, $\Lambda$, and the number of
points in the $z$-averaging) there are other systematic errors which
do not average out, thus the NRG spectra may deviate more from the
true spectra than the proposed error estimates indicate, but there is
no way to detect such effects within the NRG itself. Nevertheless, the
knowledge about the amplitude of the oscillatory discretization and
truncation errors, especially in relation with the spectral broadening
problem (discussed below), is important to make the best possible use
of the method.

\section{Model and details of the numerical technique}

We study the single-impurity Anderson model \cite{anderson1961,
hewson} $H=H_\mathrm{imp}+H_\mathrm{c}+H_\mathrm{bath}$ with
\begin{equation}
H_\mathrm{imp}=\epsilon \sum_\sigma d^\dag_\sigma d_\sigma
+ U n_\uparrow n_\downarrow + g\mu_B B (n_\uparrow-n_\downarrow)/2,
\end{equation}
while $H_\mathrm{c}$ and $H_\mathrm{bath}$ are the coupling and the
conduction band parts. The electron repulsion is $U=0.1$, the on-site
energy is $\epsilon_d=-U/2$, the hybridization is $\Gamma=0.008$, and
the external magnetic field is $b=g\mu_B B=3.273\ 10^{-4}$. All
parameters are expressed in units of the half-bandwidth $D=1$. Unless
otherwise noted, the discretization parameter is $\Lambda=2$, $N_z=12$
discretization meshes are used, and the broadening parameter is
$\alpha=0.075$ using the broadening kernel proposed in
Ref.~\onlinecite{weichselbaum2007}. (After rescaling by $D=10$ and
adopting a different convention for expressing the hybridization
strength as $\Gamma/D=0.16$, these are the same parameters as used in
Ref.~\onlinecite{schmitt2011}; in a later section, we will, as an
example, provide an approximation for the Kondo-peak splitting
together with an error estimate for this parameter set.) The spectra
are broadened and $z$-averaged before the self-energy trick is
applied. All calculations have been performed for zero temperature; in
this case the CFS method \cite{peters2006} and the full density-matrix
method \cite{weichselbaum2007} become fully equivalent. In the NRG
calculation, the same maximum number of states $N$ is kept in the
truncation for all values of $z$; this is necessary for a meaningful
$z$-averaging in the CFS approach. (If, instead, the spectral
functions are calculated using the alternative patching approach
\cite{costi1994, bulla2001}, it is advantageous to use the truncation
with a fixed energy cutoff; see also Ref.~\onlinecite{resolution},
where the systematic errors in the patching approach are
comprehensively analyzed).

\begin{figure}[htbp!]
\centering
\includegraphics[width=8cm,clip]{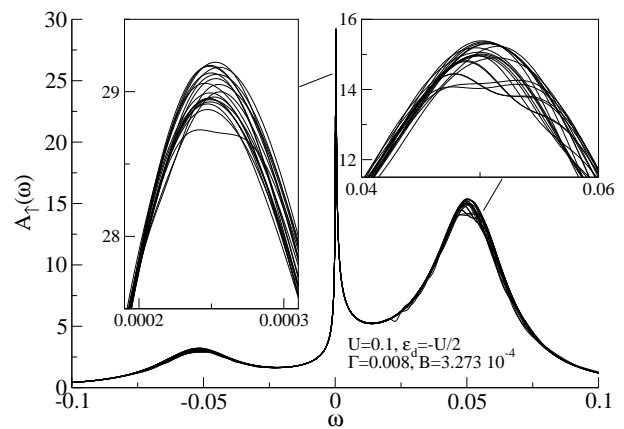}
\caption{Spectral function $A_\uparrow(\omega)$ of the Anderson
impurity model calculated for a range of $N$, the number of states
kept in the truncation after each NRG iteration step. $N$ ranges from
1800 to 3600 in steps of 100. The Kondo temperature (defined as in
Refs.~\onlinecite{wilson1975,krishna1980a}) is $T_K=6.9\ 10^{-5}$,
thus $B/T_K=4.7$.
}
\label{fig1}
\end{figure}

\begin{figure}[htbp!]
\centering
\includegraphics[width=6cm,clip]{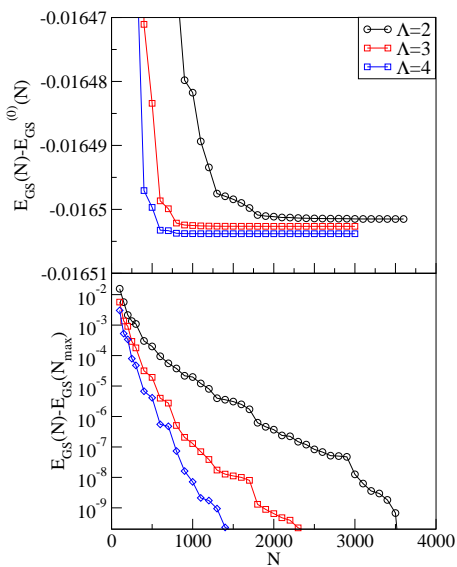}
\caption{(Color online) Ground-state energy $E_\mathrm{GS}$ as a
function of the number of states kept, $N$. $E_\mathrm{GS}$ is computed
as described in Ref.~\onlinecite{groundstate}. 
$E_\mathrm{GS}^{(0)}$ is the ground-state energy of the conduction
band without the impurity, thus what is plotted is the impurity
binding energy. The variation of the binding energy (in the large-$N$
limit) with $\Lambda$ is a direct consequence of the discretization
errors. These results indicate the degree of convergence of the NRG
calculation as a function of $N$ and $\Lambda$.}
\label{fig4}
\end{figure}

In Fig.~\ref{fig1} we plot the spectral functions obtained in the
calculations for a variable number of states kept, ranging from
$N=1800$ (the value used in Ref.~\onlinecite{schmitt2011}) up to
$N=3600$. It has to be noted that even the lower limit $N=1800$ is
sufficient to obtain essentially fully converged NRG results for
static quantities (ground state energy, thermodynamic functions); the
plot in Fig.~\ref{fig4} shows that at $N=1800$ the error in the
calculated ground state energy is below $10^{-6}$, which is already a
highly precise result. As evidenced in Fig.~\ref{fig1}, the spectral
functions, however, do not converge to some limiting curve as $N$ is
increased. This is due to the way the spectral function is calculated
in the CFS approach \cite{peters2006,weichselbaum2007}: the delta-peak
contributions appear at frequencies $\omega$ which are a difference of
the energy of a kept state $E_k$ and the energy of a discarded state
$E_d$, that is, $\omega=E_d-E_k$. While the kept states are in the
part of the on-shell excitation spectrum which is expected to be
rather accurate, thus $E_k$ is precise, the truncated states come from
the top of the spectrum which is more significantly affected by the
accumulated truncation errors from the previous NRG steps, thus the
energies $E_d$ of these states are known to a lesser precision. In
other words, in the CFS approach the normalization of the spectral
function is guaranteed to be exactly 1, but the higher moments are not
exact, i.e., only the 0-th spectral sum rule is fulfilled to numeric
precision \cite{peters2006, resolution}. Increasing $N$ does not help
in this respect, since this merely implies that a given excitation
will contribute at a later NRG step, thus $E_d$ may accumulate even
more truncation error. For this reason, changing $N$ will change the
resulting spectrum. Since the states in the NRG are clustered, the
changes can be relatively abrupt. (We note in passing that the
spectral functions calculated using the patching approach converge as
$N$ is increased, because one extracts the spectral information always
from the same energy interval of the on-shell spectra, thus the effect
of the truncation errors of the discarded states is tiny. Alas, the
patching approach suffers from other deficiencies; in particular,
there is a free parameter which needs to be tuned for each particular
application, which limits the reliability of the approach.)

\section{Estimation of the spectral-function variance}

\begin{figure}[htbp!]
\centering
\includegraphics[width=8cm,clip]{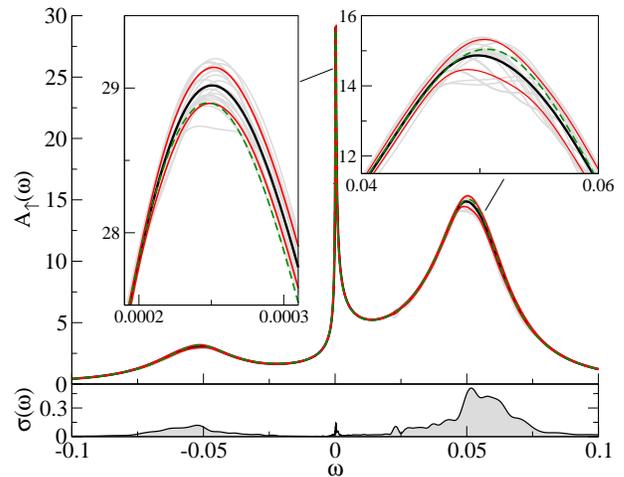}
\caption{(Color online) Upper panel: Spectral function calculated by
averaging the CFS spectra over all values of $N$ (thick black line)
with a confidence interval determined by the standard deviation of the
data at each $\omega$ (thinner lines/red online). For comparison, the
spectral function calculated using the patching approach is also shown
(dashed line/green online). The gray lines in the background are
the individual CFS spectra from Fig.~\ref{fig1}. Lower panel: standard
deviation as a function of frequency.}
\label{fig2}
\end{figure}

Fig.~\ref{fig2} shows the spectral function $A(\omega)$ obtained by
averaging over the curves shown in Fig.~\ref{fig1}, together with the
confidence region which represents the estimated range of values over
which the spectral function fluctuates as a function of the truncation
cutoff $N$. The confidence region is determined by calculating the
standard deviation $\sigma(\omega)$ at each frequency $\omega$; the
upper and lower boundaries of the region are then taken to be
$A(\omega)\pm \sigma(\omega)$. For comparison, the spectral function
calculated using the traditional patching approach is also shown; it
is evident that this curve lies within or near the confidence region
of the spectral function calculated using the CFS approach. In this
sense, the two approaches appear equivalent in this case, however the
patching approach by itself provides no means for estimating the
reliability (i.e., the confidence region) of the result. The lower
panel in Fig.~\ref{fig2} shows the standard deviation,
$\sigma(\omega)$. It attains its highest values near the atomic peaks
at $\omega \sim \epsilon_d, \epsilon_d+U$ and it tends to decrease at
lower frequencies, as expected, although it has a small local maximum
at the Kondo resonance where the results are again more scattered.

\begin{figure}[htbp!]
\centering
\includegraphics[width=2cm,clip]{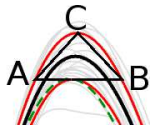}
\caption{(Color online) Estimation of the true position of the peak
maximum; see the text for details. The data is taken from the left
inset in Fig.~\ref{fig2}.}
\label{fig3}
\end{figure}

From the results in Fig.~\ref{fig2} we extract the position of the
Kondo resonance as the frequency of the maximum of the averaged
spectral function, $\omega_\mathrm{max}=2.50\ 10^{-4}$. Let us
consider now the estimation of the error committed due to the variance
of the spectral functions. As a simple (albeit pessimistic)
approximation we may consider that the true maximum is located
anywhere within the triangle plotted in Fig.~\ref{fig3}. The average
width of the triangle is $(\omega_B-\omega_A)/2$, thus the error
estimate can be defined as
\begin{equation}
\delta \omega = \frac{\omega_B-\omega_A}{4}.
\end{equation}
From the calculations for the discretization $\Lambda=2$ and the
broadening $\alpha=0.075$ we thus conclude that the Kondo resonance is
shifted by the external magnetic field to
\begin{equation}
\omega_K \approx 2.50\ 10^{-4}\ \pm\ 1.2\ 10^{-5} \approx
2.50\ 10^{-4}\ \left( 1 \pm 0.05 \right),
\end{equation}
or
\begin{equation}
\omega_K/b \approx 0.76 \left(1 \pm 0.05 \right),
\end{equation}
where $b=g\mu_B B$ is the magnetic field in energy units (Zeeman
energy). In other words, using the NRG calculations at $\Lambda=2$,
the position of the Kondo peak can be determined at best with 5\%
accuracy. This is the reason why it is so difficult to reliably study
the dependence between the Kondo peak position and the external
magnetic field, which should behave as $\omega_K = (2/3) b$ at low
fields ($b \ll k_B T_K$) \cite{logan2001,hewson2006} and as $\omega_K
= b$ at high fields \cite{costi2000, moore2000,
konik2001,rosch2003,hewson2006,meir1993} ($b \gg k_B T_K$, but $B$
still small compared to the atomic parameters of the model, otherwise
non-universal features are observed \cite{aniso2,schmitt2011}).

\begin{figure}[htbp!]
\centering
\includegraphics[width=8cm,clip]{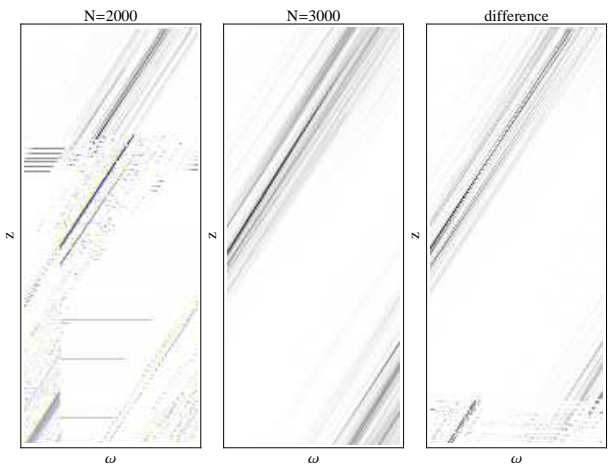}
\caption{Raw binned spectral data in the energy range of the Kondo
peak for a fine-grained range of $z$ values ($N_z=256$). These are raw
spectral weights of the delta peaks forming the spectral functions
which have been binned (logarithmic mesh, $600$ bins per decade; the
intervals of the binning grid are much narrower than the broadening
kernels used to post-processes these raw results). Left and middle
panels show results for $N=2000$ and $N=3000$, while the right panel
shows the difference between the two. The grayscale ranges are
approximately equal in the three plots. In the plots $z$ ranges from 0
to 1 (bottom to top) and $\omega$ ranges from $0.0002$ to $0.0003$
(across the Kondo peak). The model parameters are as in
Fig.~\ref{fig1}. }
\label{comb1}
\end{figure}

To gain more insight into the origin of the variation of the spectral
function with $N$, in Fig.~\ref{comb1} we plot the raw binned spectral
data for a fine-grained set of 256 different $z$ values. A comparison
of the results for $N=2000$ and $N=3000$ shows that while the general
aspects do not vary with $N$, the details do. The differences in these
details lead to the variation of the spectral function with increasing
$N$.

\section{Spectral function broadening}

We now discuss the role of the spectral function broadening parameter
$\alpha$ which is used to obtain smooth spectral curves from the
spectral information in the form of weighted delta peaks. The
broadening kernel used is
\begin{equation}
P(\omega,\omega') = \frac{\theta(\omega \omega')}{\sqrt{\pi} \alpha
|\omega|} \exp\left[ -\left(\frac{\log|\omega/\omega'|}{\alpha}
-\frac{\alpha}{4} \right)^2 \right],
\end{equation}
i.e., the broadening kernel proposed in
Ref.~\onlinecite{weichselbaum2007} with $\gamma=\alpha/4$. The peak of
this kernel function is located at
\begin{equation}
\omega=e^{-\alpha^2/4} \omega'.
\label{shift}
\end{equation}
The weight is distributed asymmetrically with respect to $\omega'$
with $1/2[1+\mathrm{erf}(\alpha/4)]$ of the weight in the interval
$|\omega|>|\omega'|$ and $1/2[1-\mathrm{erf}(\alpha/4)]$ in the
interval $|\omega|<|\omega'|$, where $\mathrm{erf}$ is the error
function. Furthermore, the tail of the
broadening function reaches to relatively high frequencies if $\alpha$
is large, thus the presence of high-energy spectral features might
lead to spurious shifts of the peak positions in the low-energy part
of the spectrum, and vice versa.

\begin{figure}[htbp!]
\centering
\includegraphics[width=8cm,clip]{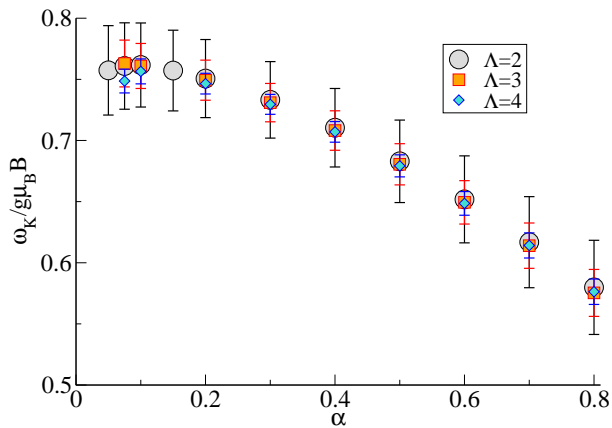}
\caption{ (Color online) The Kondo peak position (with ``pessimistic''
error bars) as a function of the broadening parameter. The (average)
peak position is extracted by determining the maximum of the average
of the spectral functions for different $N$; the maximum is located
using the conjugate gradient method. The smallest $\alpha$, for which
the Kondo peak maximum can still be extracted, is determined by the
number $N_z$ of different values of the twist parameter (here
$N_z=12$, and $\alpha_\mathrm{min} \approx 0.05$). For completeness,
the results for larger broadening parameters $\Lambda=3$ and
$\Lambda=4$ are also shown; smaller error-bars result from the fact
that for coarser broadening the truncation errors decrease (while the
discretization errors increase). Thus larger values of $\Lambda$
actually lead to less scatter in the calculated spectral functions as
$N$ is varied.}
\label{fig5}
\end{figure}

Naively, one could expect that increasing $\alpha$ will reduce the
variance of the spectral curves. This is only partially true: the
irregularities indeed smooth out, however the overall spectral weight
distribution is still found to be fluctuating with $N$. Surprisingly,
it is found that that $\delta \omega_K$, the estimated (pessimistic)
error in the Kondo peak position, remains roughly constant with
$\alpha$, see Fig.~\ref{fig5} for $\Lambda=2$. It is thus important to
use a small broadening parameter to avoid systematic overbroadening
errors \cite{schmitt2011, zhang2010kondo}. Clearly, a suitably small
$\alpha$ is such that the overbroadening error is smaller than the
intrinsic error due to the discretization (for instance, no larger
than $\alpha=0.2$ or $\alpha=0.3$ in our example, see
Fig.~\ref{fig5}). The broadened spectral functions depend on the
broadening kernel used; only in the limit of very small broadening
widths do the different broadening kernels (Gaussian, log-Gaussian,
modified log-Gaussian) all become equivalent. For moderate and large
broadening, it was found that the modified log-Gaussian kernel works
best, see Fig.~19 in Ref.~\onlinecite{resolution}.

It may be noted that the variance does not depend on the choice of the
discretization scheme, because for the chosen model parameters and
$\Lambda=2$, the discretization artifacts are small in any
discretization scheme.

\begin{figure}[htbp!]
\centering
\includegraphics[width=8cm,clip]{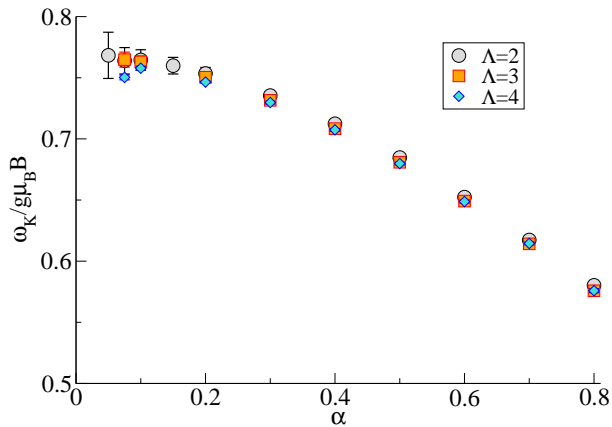}
\caption{(Color online) The Kondo peak position (with ``optimistic''
error bars) as a function of the broadening parameter. The peak
position is extracted as an average of the maxima of the spectral
functions for different $N$; each maximum is located using the
conjugate gradient method. The error bars are determined by
calculating the standard deviation of the extracted values for
$\omega_K$. The error bars increase significantly for $\alpha < 0.1$
because at given number of the twist parameters, $N_z=12$, the
oscillations in the spectral functions become severe for $\alpha
\lesssim 1/\sqrt{\Lambda}N_z$.}
\label{optimistic}
\end{figure}

Alternatively, one can determine the error in the extracted Kondo peak
position by locating the maximum for each spectral function calculated
at fixed $N$, and calculating the standard deviation. The results of
such a calculation are shown in Fig.~\ref{optimistic}. The error bars
are significantly smaller using this definition and they decrease with
increasing $\alpha$. We can describe this procedure as ``optimistic'',
since it is found that the spectral functions with different $N$ in
fact have their maxima at positions which fluctuate less that the
curve does overall, thus the error bars are significantly smaller than
in the ``pessimistic'' case presented in Fig.~\ref{fig5}. The true
systematic error probably lies somewhere between these two extreme
error estimates, thus a suitable value of $\alpha$ for performing
calculations aiming towards high-precision results is smaller than
suggested by the error bars in Fig.~\ref{fig5}, i.e., the broadening
parameter should probably be closer to $\alpha=0.1$, rather than
$\alpha=0.2$ or $0.3$, as suggested above.

It is interesting to note that the effect of finite $\alpha$ is larger
than expected from the shift described by Eq.~\eqref{shift}. A good
fit to the results for $\Lambda=2$ in Figs.~\ref{fig5} and
\ref{optimistic} is 
\begin{equation}
\omega_K(\alpha) = \omega_K(0) e^{-\alpha^2/b}
\end{equation}
with $b \approx 2.25$. Thus the functional form of the shift is
similar to that of Eq.~\eqref{shift}, however the numerical factor in
the argument of the exponential function is $2.25$ rather than $4$,
i.e., the shift is even larger than expected. Test calculations for a
single Lorentzian peak indeed show that spectral peaks with a finite
width are broadened into a wider peak whose maximum is displaced by a
factor of $e^{-\alpha^2/b}$ with $b$ smaller than 4. This is another
reason for reducing $\alpha$ as far as possible.

\section{Kondo resonance splitting}

\begin{figure}[htbp!]
\centering
\includegraphics[width=8cm,clip]{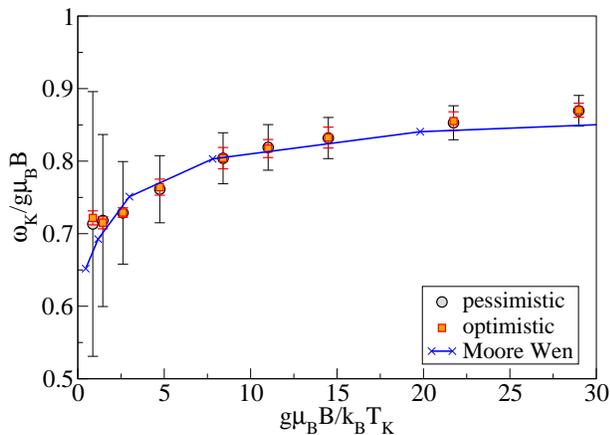}
\caption{(Color online) The Kondo resonance splitting as a function of
the external magnetic field. The error bars correspond to the
"pessimistic" and "optimistic" error estimates. Model parameters are
$U=0.1$, $\epsilon_d=-U/2$, $\Gamma=0.008$ in units of half-bandwidth
of the (flat) conduction band. The broadening parameter is
$\alpha=0.075$. The data labeled as ``Moore Wen'' are taken from
Fig.~2 in Ref.~\onlinecite{moore2000}.}
\label{figsplit}
\end{figure}

As an application, we now consider the problem of the Kondo resonance
splitting in an external magnetic field. We plot the splitting ratio
$\omega_K/(g\mu_B B)$ as a function of the ratio $g\mu_B B/k_B T_K$ in
Fig.~\ref{figsplit}. (Note that $\omega_K$ is defined as the shift of
the Kondo peak position in the spin-projected spectral function
$A_\sigma(\omega)$ and that the peak-to-peak distance in the
spin-averaged spectral function is not exactly $2\omega_K$.) Both
procedures for extracting the average value and the error bars have
been performed. The average values (circles and squares in the figure)
agree within the error bars of the "optimistic procedure" for all the
results shown; deviation becomes larger for smaller magnetic fields.
For large fields, the ratio increases in a rather slow (logarithmic)
way, thus one expects non-universal features to appear before the
universal high-field asymptotic behavior is reached. For small fields,
the ratio goes toward a value of 0.7, which is close to the expected
low-field limit value \cite{logan2001,hewson2006} of 2/3. [The
low-field asymptotic limit has been reported to be confirmed in a
calculation where broadening is performed using Lorentzian peaks with
constant width at very low energy scales \cite{hewson2006}.] The
"pessimistic" error bars grow larger for small fields, and the
"optimistic" average values start to deviate from the "pessimistic"
ones. This is expected, since the Kondo peak displacement becomes
smaller than the Kondo peak width, thus the relative errors (i.e., the
error of the ratio $\omega_K/g \mu_B B$) grow with decreasing $B$
because the absolute error $\delta \omega_K$ approximately saturates
for $g\mu_B B \ll k_B T_K$.

The results in Fig.~\ref{figsplit} are in good agreement with the
splitting in the Kondo model as determined from the spinon density of
states in the Bethe Ansatz (BA) solution \cite{moore2000}, see the
full line with crosses in Fig.~\ref{figsplit}. At large fields, the
deviation beyond the error bars is most likely due to the differences
between Anderson and Kondo models at high fields which leads to
non-universal features. At low fields, we can say, at most, that the
NRG and BA results are consistent within the errorbars. This trend is
also found in experimental results \cite{quay2007}. In experiments
where the splitting was found to exceed the predicted splitting in the
high-field range \cite{kogan2004,liu2009split}, this is likely due to
the non-universal effects which are expected in systems which are not
in the extreme Kondo limit, i.e., the regime where the Kondo
temperature is lower by many orders of the magnitude compared to all
other energy scales in the problem. Such non-universal effects are
expected in the Anderson model \cite{schmitt2011}, but also in the
Kondo model \cite{aniso2}. Furthermore, one should take into account
the strong asymmetry of the Kondo peaks in strong magnetic field:
perturbative renormalization group calculations \cite{rosch2003} show
that the maximum of the spectral peak is located at $\omega > g\mu_B
B$, while the center of the left flank of the peak appears to be
position almost precisely at $\omega=g\mu_B B$. For very large fields,
the peak itself is no longer observable and one is left with a step in
the spectral function, which should be considered as the sole remnant
of the Kondo resonance. Finally, it should also be noted that for
meaningful comparison with the experimental results, it is necessary
to take into account the non-equilibrium effects if the splitting is
extracted from the conductance at finite source-drain voltage in
quantum-dot setups with symmetric coupling to both leads
\cite{hewson2005b,schmitt2011noneq}.

We are now in the position to critically discuss the recent work on
the scaling of the magnetic-field-induced Kondo resonance splitting
\cite{zhang2010kondo}, the comment concerning that work
\cite{schmitt2011}, and the reply offered by the authors of the
original work \cite{zhang2011}. In particular, it has been claimed
\cite{zhang2010kondo} that the position of the Kondo resonance in the
total spectral function does not approach its position in the
spin-resolved spectra for high magnetic fields, in contradiction to
what has been found in some previous works \cite{aniso2}, and that the
splitting shows non-universal behavior even for modest $B/T_K$ ratio
of order 10. Both conclusions have been shown in
Ref.~\onlinecite{schmitt2011} to be a consequence of the spectral
function overbroadening due to an excessively large broadening
parameter $\alpha=0.8$ and it was pointed out that different results
are obtained with smaller broadening $\alpha=0.075$. In reply, it has
been claimed that the value of $\alpha=0.075$ is too small and that
$\alpha=0.4$ is a more appropriate choice \cite{zhang2011}. 

The results of the present work, in particular Figs.~\ref{fig5} and
\ref{optimistic}, make it possible to go beyond the purported
``certain arbitrariness'' in the choice parameters \cite{zhang2011}
and elucidate to what extend the NRG can provide a definitive answer
to the problem of the magnetic-field-induced Kondo peak splitting. Two
points need to be emphasized: i) while $\alpha=0.8$ is clearly too
large (it leads to an error in excess of 30\% in determining the Kondo
peak positions) and $\alpha=0.4$ is better (15\% error), it is crucial
to go in the $\alpha \to 0$ limit in order to obtain a result with
accuracy in the percent range, thus $\alpha=0.075$ is a good choice
for $N_z=12$; ii) using the ``pessimistic'' error estimate, there is a
sizable overlap of the confidence regions for $\alpha=0.075$ and
$\alpha=0.4$, thus there is non-negligible possibility that a
calculation performed for a certain fixed value of $N$ would yield
similar results for the peak position using both values of $\alpha$,
see Fig.~\ref{fig5}, although this conclusion is likely to be too
pessimistic and a different error estimate suggest a clear difference
between $\alpha=0.4$ and $\alpha=0.075$ results, see
Fig.~\ref{optimistic}. It may be thus concluded that $\alpha=0.075$ is
a more appropriate choice of the broadening parameter and that the
results and conclusions of Ref.~\onlinecite{zhang2010kondo} are
questionable due to spectral overbroadening. In particular, in the
$\alpha \to 0$ limit, the Kondo peak positions in the total and
spin-resolved spectral functions approach in the high-field limit
\cite{aniso2,schmitt2011}. The conclusion of
Ref.~\onlinecite{zhang2010kondo} that the slope coefficient of the
ratio of the Kondo peak splitting over magnetic field is 2/3 in the
small field limit, as expected \cite{logan2001,hewson2006}, is
surprising given the significant shift of the Kondo peak due to
overbroadening; this result may be simply fortuitous. Using
log-Gaussian broadening and taking into account the error bars in the
small-$\alpha$ limit, such slope determination cannot be made in a
reliable way using the NRG, see Fig.~\ref{figsplit}. Furthermore, it
may also be remarked that the number of states kept in
Ref.~\onlinecite{zhang2010kondo}, i.e. $N=150$, is too small to obtain
well converged results (see Fig.~\ref{fig4}), irrespective of the
broadening procedure used.

\section{Inelastic (spin-flip) tunneling step and the large-field
limit}

\begin{figure}[htbp!]
\centering
\includegraphics[width=8cm,clip]{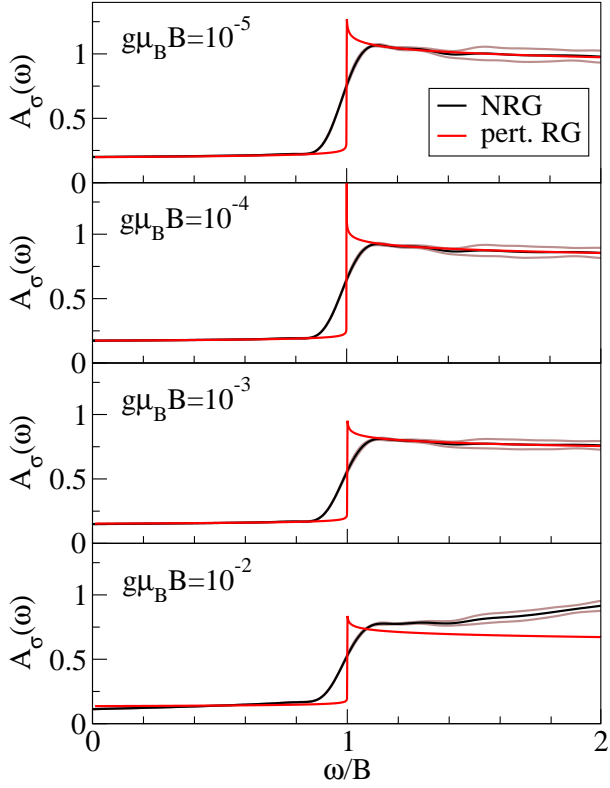}
\caption{(Color online) Spectral function of the Anderson impurity
model in high magnetic field, $B \gg T_K$. Black line corresponds to
the NRG results, the two ligher (brown onlines) lines demarkate the
confidence region, while the light (red online) curve corresponds to
the analytical perturbative RG function from
Ref.~\onlinecite{rosch2003}.}
\label{figB1}
\end{figure}

If $U$ is much larger than $\Gamma$, for instance $U/\Gamma=100$, the
Kondo temperature is for all practical purposes equal to zero, since
experiments are performed at a finite temperature which is, in this
case, larger than $T_K$ by orders of magnitude. The impurity then
behaves much like a free spin, as long as the external magnetic field
is not comparable to the atomic energy scales ($\epsilon$, $U$). In
experiments, for example in the inelastic spin-flip tunneling
spectroscopy using a scanning tunneling miscroscope (STM)
\cite{heinrich2004}, one can induce inelastic scattering by injecting
electrons from the STM tip into the adsorbed impurity with energy
exceeding the characteristic energy of a spin-flip event, i.e., above
the Zeeman energy. Due to high relevance for STM experiments, it is of
substantial interest to study the spectral function of the Anderson
impurity model in the vicinity of the on-set of inelastic scattering.
We study three aspects of this problem: i) the line-shape of the step
in the spectral function at the on-set of spin-flip scattering; ii)
the position of this step as $B$ approaches the atomic scales; iii)
merging of the step with the atomic peak.

\begin{figure}[htbp!]
\centering
\includegraphics[width=8cm,clip]{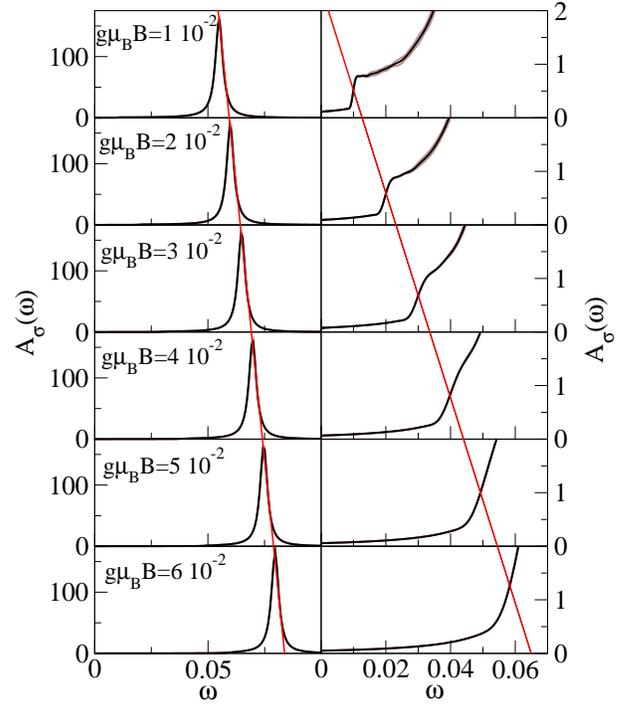}
\caption{(Color online) Spectral function of the Anderson impurity
model in very high magnetic field, $B \gg T_K$ and $B \sim U/2$. The
dark lines are the averaged NRG results, while the lighter (brown
online) lines indicate the confidence region. The discretization
errors are more pronounced on the high-frequency side of the spectral
step at $\omega=b \equiv g\mu_B B$.}
\label{figB2}
\end{figure}

We express the magnetic field in energy units (Zeeman splitting) as
$b=g\mu_B B$. For $b$ small compared with the atomic scales, but $b
\gg T_K$, we find that the spectral function around $\omega=b$ takes
the form of a step, see Fig.~\ref{figB1}. This regime has been studied
before using both NRG and perturbative renormalization group (RG)
techniques, see Fig.~5 in Ref.~\onlinecite{rosch2003}. Our results are
consistent with these studies. Taking into account the finite
broadening width in the NRG calculations, an excellent agreement is
found with perturbative RG as long as $b$ is much smaller compared to
$U$ (see the deviations for $b/U = 1/5$, bottom-most panel in
Fig.~\ref{figB1}; note that the perturbative RG calculation is
performed for the effective Kondo model, not for the Anderson model).
In experiments, finite temperature will play a similar smoothing
effect as spectral broadening in NRG, thus one can indeed expect to
observe a step-like spectral line-shape. The step, whose center is
always located at $\omega=b$, can be interpreted as the on-set of the
inelastic (spin-flip) scattering, which is observed by the spin-flip
spectroscopy in system that do not exhibit the Kondo effect
\cite{heinrich2004}. It should be noted that there is no discernable
peak at $\omega=b$: in this $b \gg T_K$ regime, the field-split Kondo
resonance has become so asymmetric that it takes the form of a
relatively sharp step \cite{rosch2003}. The width of the step as
determined by the NRG matches that expected for a unit-step function
broadened by the kernel, i.e., the intrinsic width of the step is very
small, presumably equal to $\Gamma \approx \pi B/(16\ln^2[B/T_K])$,
the transverse spin relaxation rate \cite{rosch2003}. We also point
out that nothing noticeable happens on the scale $b \sim
\Gamma=10^{-3}$, nor on the scale $b \sim \sqrt{U\Gamma} = 10^{-2}$. 

We now study the regime where $b$ is comparable to the atomic scales,
Fig.~\ref{figB2}. We again observe that there is always a spectral
step exactly at $\omega=B$, see right panels in Fig.~\ref{figB1}. The
steps becomes more diffuse as it merges with the atomic peak for $b
\gtrsim U/2$.

In the atomic limit ($\Gamma\to0$), the impurity energy levels are 
\begin{equation}
\begin{split}
E_0 &= 0,\\
E_\uparrow &= \epsilon+b/2,\\
E_\downarrow &= \epsilon-b/2,\\
E_2 &= 2\epsilon+U.
\end{split}
\end{equation}
In the particle-hole symmetric point, $\delta=\epsilon+U/2=0$, one has
$E_2=E_0$, thus for any non-zero value of the magnetic field, the
state $\ket{\downarrow}$ is the ground state and the spin-up spectral
function has a peak at
\begin{equation}
\omega_0=E_2-E_\downarrow=\epsilon+U+b/2=\delta+U/2+b/2.
\end{equation}
We find that for $b \gtrsim U$, the peak position indeed deviates only
little from $\omega_0$, see left panels in Fig.~\ref{figB2}.

\section{Conclusion}

An analysis of the spectral functions calculated using the NRG
technique shows that there is always some variance due to the
discretization and truncation errors. For the range of values of $N$
which is suitable for practical NRG calculations, the obtained
spectral functions do not converge; instead, the variance of the
truncation errors appears to be approximately constant as a function
of $N$ (for large $N$). Using a large broadening parameter does not
solve the problem, but merely masks it. Furthermore, overbroadening
errors appear to be a much larger reason for concern than the
discretization and truncation errors. Accurate calculations should
therefore aim for obtaining the results in the limit of zero
broadening width, taking into account the constraints imposed by the
systematic NRG errors. Errors should be quantified, not ignored.

We applied the procedure to estimate the errors in the NRG results for
the Kondo resonance splitting in the external magnetic field. The
systematic errors preclude the study of the small-field limit. For
intermediate fields, however, it is possible to calculate the
splitting ratio $\omega_K/g \mu_B B$ with an estimated error of a few
percent, which is reasonably accurate. Good agreement is found with
the Bethe Ansatz results for the peak splitting by Moore and Wen
\cite{moore2000} in the regime of small and intermediate magnetic
fields, where Anderson and Kondo models are equivalent.

To find a definitive quantitative solution of the problem of the Kondo
resonance splitting in the external magnetic field, it will be
necessary to devise numerical techniques that can reduce the
discretization and truncation artifacts even further. Recent
developments where the NRG matrix product state representation is
refined using the density-matrix renormalization group (DMRG)
procedure or variationally \cite{weichselbaum2009,pizorn2011} may be
very valuable in this respect since they might allow performing
calculation at much reduced discretization parameters $\Lambda$ and
optimizing the excited state energies by sweeping.

\begin{acknowledgments}
I thank I. Pi{\v z}orn, S. Schmitt, J. Bauer, Th. Pruschke and R.
Peters for discussions and acknowledge the support of the Slovenian
Research Agency (ARRS) under Grant No. Z1-2058.
\end{acknowledgments}

\bibliography{errors}

\end{document}